\documentclass[conference,compsoc]{IEEEtran}
\usepackage{amssymb}
\usepackage{pifont}
\newcommand{\cmark}{\ding{51}}%
\ifCLASSOPTIONcompsoc
  \usepackage[nocompress]{cite}
\else
  \usepackage{cite}
\fi
\ifCLASSINFOpdf
  \usepackage[pdftex]{graphicx}
  \usepackage{graphicx}
  \graphicspath{ {./images/} }
  \usepackage[outdir=./images/]{epstopdf}
\else
  \usepackage[pdftex]{graphicx}
  \usepackage{graphicx}
  \graphicspath{ {./images/} }
  \usepackage[outdir=./images/]{epstopdf}
\fi
\usepackage{array}
\usepackage[flushleft]{threeparttable}
\newcolumntype{P}[1]{>{\centering\arraybackslash}p{#1}}

\usepackage{stfloats}
\usepackage{url}
\usepackage{doi}

\begin{document}
%
\title{Behind The Wings: The Case of Reverse Engineering and Drone Hijacking in DJI Enhanced Wi-Fi Protocol}


\author{\IEEEauthorblockN{Derry Pratama\IEEEauthorrefmark{1}\IEEEauthorrefmark{3},
Jaegeun Moon\IEEEauthorrefmark{2}\IEEEauthorrefmark{4},
Agus Mahardika Ari Laksmono\IEEEauthorrefmark{1},\\ 
Dongwook Yun\IEEEauthorrefmark{1},
Iqbal Muhammad\IEEEauthorrefmark{1},
Byeonguk Jeong\IEEEauthorrefmark{1},
Janghyun Ji\IEEEauthorrefmark{2} and
Howon Kim\IEEEauthorrefmark{1}\IEEEauthorrefmark{5}
}
\IEEEauthorblockA{\IEEEauthorrefmark{1}School of Electrical and Computer Engineering\\
Pusan National University, Busan, South Korea\\
\IEEEauthorrefmark{3}Email: derryprata@gmail.com\\
\IEEEauthorrefmark{5}Email: howonkim@pusan.ac.kr}
\IEEEauthorblockA{\IEEEauthorrefmark{2}SmartM2M, South Korea\\
\IEEEauthorrefmark{4}Email: jaekun34@smartm2m.co.kr}}


%


\maketitle

\begin{abstract}
This research paper entails an examination of the Enhanced Wi-Fi protocol, focusing on its control command reverse-engineering analysis and subsequent demonstration of a hijacking attack. Our investigation discovered vulnerabilities in the Enhanced Wi-Fi control commands, rendering them susceptible to hijacking attacks. Notably, the study established that even readily available and cost-effective commercial off-the-shelf Wi-Fi routers could be leveraged as effective tools for executing such attacks. To illustrate this vulnerability, a proof-of-concept remote hijacking attack was carried out on a DJI Mini SE drone, whereby we intercepted the control commands to manipulate the drone's flight trajectory. The findings of this research emphasize the critical necessity of implementing robust security measures to safeguard unmanned aerial vehicles against potential hijacking threats. Considering that civilian drones are now used as war weapons, the study underscores the urgent need for further exploration and advancement in the domain of civilian drone security.
\end{abstract}


%
\IEEEpeerreviewmaketitle

\section{Introduction}
Drones have become increasingly popular in recent years and are widely used for both industrial and personal purposes. Drones were expensive back then and only used by experts for specific missions. But nowadays, the price is getting cheaper; even a toy drone can be purchased from the range of \$10 to \$1000. This price range affects the drone's features, its development cost, and of course, its security. Due to constrained hardware and development costs, many consumer drones are sold cheaply and built with poor security. The lack of security verification and encryption makes them vulnerable to different kinds of attacks such as replay attack~\cite{ref_replyattack} or even GPS spoofing~\cite{ref_gpsspoofing1}\cite{ref_gpsattack1}.

The discovery of vulnerabilities and associated attacks on civilian drones may not yield substantial consequences when targeting drones solely employed for recreational purposes or video recording. However, the landscape has evolved, and civilian drones are now being utilized as weapons, especially to carry explosives and kamikaze drones in the ongoing Ukrainian and Russian conflicts~\cite{ref_budget_drone_warfare}. Ukraine's Defense Ministry recognized the power of small drones early in the conflict and appealed to the owners of small drones to use them for missions in Kyiv~\cite{ref_DroneMilitary}. Consumer drones are easy to get, they can be easily modified to carry a grenade or other small explosive, which can be dropped with great precision into trenches filled with troops or directly into the open top of a tank.~\cite{ref_chinesedronehobby}~\cite{ref_dronevstank}

This raises the question of how secure civilian drones currently are. Research has been conducted on drone communication, as discussed in ~\cite{ref_dronesecurity}, which surveys drone security and previous attempts of drones attack using different methods on different drones. However, there is still a lack of research investigating the market leader of consumer drones, DJI, which is used by most people. Recent research~\cite{ref_droneid} found that DJI exposed the DroneID in the Ocusync protocol unencrypted, containing the pilot's location. This raises questions about the security awareness of DJI drones, especially about the other protocol. Other DJI drones, such as Mini SE, communicate using Enhanced Wi-Fi which is, which comes from Wi-Fi protocol that has already vulnerable to some kinds of authentication~\cite{ref_kracken1} and decryption attacks~\cite{ref_wep}~\cite{ref_wpa2}.


Using consumer drones as weapons poses an immense and alarming risk that extends far beyond mere concerns about privacy. The potential consequences become apparent when weaponized consumer drones fall into the wrong hands and are exploited due to an inadequate understanding of their security measures. The urgency to address this issue cannot be overstated, as the safety and well-being of individuals are undeniably at stake. This paper aims to shed light on the security of the Enhanced Wi-Fi protocol employed by DJI drones and its vulnerability to hijacking attacks. We comprehensively analyze DJI control commands in Enhanced Wi-Fi protocol and showcase a hijacking attack utilizing a low-cost SDR and Wi-Fi router. By undertaking this research, we intend to deepen our understanding of consumer drone security and provide valuable insights into the necessary measures for safeguarding unmanned aerial vehicles against hijacking attacks.
Our contributions presented in this paper includes the following:
\begin{itemize}
\item A security analysis on the Enhanced Wi-Fi connection link and drone control implementation.
\item We prove that the current DJI Enhanced Wi-Fi protocol link is not secure enough with a proof of concept of a real-world attack to hijack the DJI Mini SE passively in fully remote conditions.
\item We also open-sourced our proof-of-concept attack code at Github repository\footnote{Github Repository: \href{https://github.com/ibndias/dji-drone-hijacking}{https://github.com/ibndias/dji-drone-hijacking}}  to allow the broader community aware of this potential risk.
\item We discuss countermeasures and the possibility of other drone communication types, and we also highlight the limitations of relying solely on Wi-Fi technology for safeguarding against hijacking attacks.
\end{itemize}
Finally, we conclude that the current market-leading drone company implementation is not secure enough, and a low-cost Wi-Fi router is enough to break it.

\section{Preliminaries}
\subsection{Drone Communication Types}
Several communications are used between the remote controller (RC) and the drone. This communication controls drone movement, exchange of information, video transmission, and status. One of the most common communication protocols used in drones is the Wi-Fi protocol, followed by a Radio Frequency, and the short-range drone sometimes uses Bluetooth.
To provide location and positioning information, the drone usually has a GNSS protocol~\cite{ref_gnss} based on a satellite navigation system. A cellular communication protocol is also used in some drones for long-range communication and data transmission, and there is the Zigbee protocol which provides low-power communication protocol for drone-to-drone communication; this is used to exchange data to achieve swarm behavior.
\subsubsection{Enhanced Wi-Fi Protocol}

DJI is a prominent consumer drone manufacturer renowned for its diverse range of models offering various features. Notably, certain low-end drone offerings are equipped with an Enhanced Wi-Fi feature~\cite{ref_DJIMiniSpec}, providing users the ability to select their desired operating frequency. However, the level of security afforded by this Enhanced Wi-Fi protocol remains uncertain due to limited research in this domain. Further investigation is necessary to ascertain its adequacy for safe and reliable usage.

\subsubsection{DJI Universal Markup Language}
In the community of DJI hobbyists, the term commonly utilized to refer to DJI's proprietary communication protocol is "DUML" an acronym standing for DJI Universal Markup Language. Although the official name of the protocol used by DJI remains undisclosed, some members of the community have undertaken reverse engineering efforts to gain insights into certain aspects of this communication protocol.

\subsection{Wi-Fi Protocol Security}
Wired Equivalent Privacy WEP is the oldest encryption standard and is considered insecure due to its vulnerability. It uses a 64-bit or 128-bit key to encrypt the data. Nowadays, it is no longer recommended to use since cracking takes just a little time needed. Wi-Fi Protected Access (WPA) is developed to replace WEP with a more secure encryption standard. It has a TKIP temporal key integrity protocol to encrypt data transmission and give stronger protection than WEP.

WPA2 uses the Advance Encryption Standard and is still considered secure. However, recent research found a flaw in the handshake that caused the attacker to see the encrypted message as a man-in-the-middle attack~\cite{ref_kracken1}. Fixing the vulnerability on every device is difficult, and many obsolete devices are left unpatched in the wild. WPA3 provide enhanced security features with stronger passwords, forward secrecy, and better encryption. However, the same researcher~\cite{ref_dragonblood} also found flaws in this protocol. Hardware compatibility issues, implementation challenges, development cost to upgrade, and interoperability of WPA3 are limited. Therefore, it’s not widely adopted in current devices.

\subsection{Wi-Fi Frequencies and Channel Width}
Wi-Fi operates in two primary frequency bands: 2.4 GHz and 5 GHz. The 2.4 GHz band, being one of the earliest utilized for Wi-Fi, offers good coverage and obstacle penetration but is prone to congestion and interference due to its popularity and coexistence with other wireless devices.

There are total of 14 channels, with only three non-overlapping channels (1, 6, and 11). In contrast, the 5 GHz band offers wider channel bandwidth and less congestion, resulting in higher data rates and improved performance. Although it has a more extensive range of channels, the coverage is limited compared to the 2.4 GHz band.

Wi-Fi channel width refers to the range of frequencies used to transmit data within a wireless network. While 5 MHz channel width was used in older Wi-Fi standards like 802.11a, it is essential to clarify that this channel width is no longer common in modern Wi-Fi deployments. Instead, the standard channel widths used today are 20 MHz, 40 MHz, 80 MHz, and even 160 MHz in some cases.

The 5 MHz channel width was limited in data transmission capacity and was mostly used in early Wi-Fi deployments with lower data rate requirements. As Wi-Fi technology evolved and the demand for higher data rates increased, wider channel widths became necessary to accommodate the higher data throughput.

Today, Wi-Fi devices support broader channel widths like 20 MHz, 40 MHz, and above, which enable faster data rates and improved performance. Wider channel widths allow more data to be transmitted simultaneously, leading to higher overall throughput and reduced latency in wireless networks.

While 5 MHz channel width was used in the past, it is no longer a standard channel width in modern Wi-Fi deployments. Instead, wider channel widths like 20 MHz, 40 MHz, 80 MHz, and 160 MHz are more commonly used to meet the demands of today's high-speed wireless communications.

\subsection{HackRF}
HackRF is an open-source Software-Defined Radio(SDR)~\cite{ref_edu} platform device that allows implementing a wireless communication system at a low price in the range of SDR about \$300. SDR can be controlled through a software program without changing the hardware configuration when transmitting Radio Frequency (RF).

HackRF can collect information at sample rates from 2 Msps up to 20 Msps(samples per second) within the frequency range of 1 MHz to 6 MHz, which can be utilized to analyze and manipulate various wireless communication protocols (Wi-Fi, Bluetooth, GPS, etc.)

GNU Radio, a popular open-source SDR software, is used to control HackRF, providing the ability to collect, analyze, and generate data using blocks and the Python programming language. Together with GNU Radio, HackRF can also be used as a transceiver for the IEEE 802.11 a/g/p protocol~\cite{ref_Wi-Fi}.

However, the main drawback of HackRF is that it only supports half-duplex, meaning that data cannot be transmitted and received simultaneously. Previous studies~\cite{ref_hackingkeyless} uses HackRF for replay attack that collects the communication frequency between the vehicle and the remote key, while \cite{ref_hijackingGPS} use it for GPS spoofing attack on drones.



\subsection{OpenWRT\label{section:openwrt}}
OpenWRT is an open-source router firmware that supports a wide range of devices, including older devices. While other custom firmware like DD-WRT is also available, OpenWRT is the most popular one among researchers due to its versatility and wide range of supported hardware. This paper uses OpenWRT as a tool to utilize low-cost Wi-Fi devices that support uncommon channel bandwidth, such as 5 Mhz~\cite{ref_openwrt}, which can be useful for research purposes.

\subsection{MikroTik LDF 5}
The MikroTik RBLDF-5nD (LDF 5)~\cite{ref_ldf5} is a compact and lightweight wireless system specifically designed for outdoor applications, and this unit serves as a fully integrated device that allows for point-to-point or point-to-multipoint connections. Utilizing a 5GHz frequency and equipped with a 21dBi dual-chain antenna, the RBLDF-5nD delivers robust performance in various outdoor settings.

It is commonly installed on high-gain antennas for long-distance links or utilized as a CPE (Customer Premises Equipment) to expand network reach. The system is powered by RouterOS, a robust network operating system developed by MikroTik. This platform provides an extensive array of functionalities, including, but not limited to, routing, firewall implementation, and bandwidth management. Notably, it can be customized to operate with OpenWRT~\ref{section:openwrt}, enhancing its flexibility. With a cost-effective price point of only around \$45, these advanced capabilities render it an appropriate choice for conducting research in the network security domain.

\subsection{Scapy\label{scapy}}
Scapy~\cite{ref_scapy} is a powerful and versatile Python library used for interacting with computer networks, creating and manipulating network packets, and performing various network-related tasks. It was developed by Philippe Biondi and is an open-source tool that provides a user-friendly interface to construct, dissect, send, and receive network packets. Scapy is particularly popular among network engineers, security professionals, and developers for its ability to handle a wide range of network protocols and its flexibility in crafting custom packets for testing and analysis.

\section{Security Analysis}
\subsection{Threat Model and Attack Scenario}
\begin{figure}[h]
    \includegraphics[width= 0.48\textwidth] {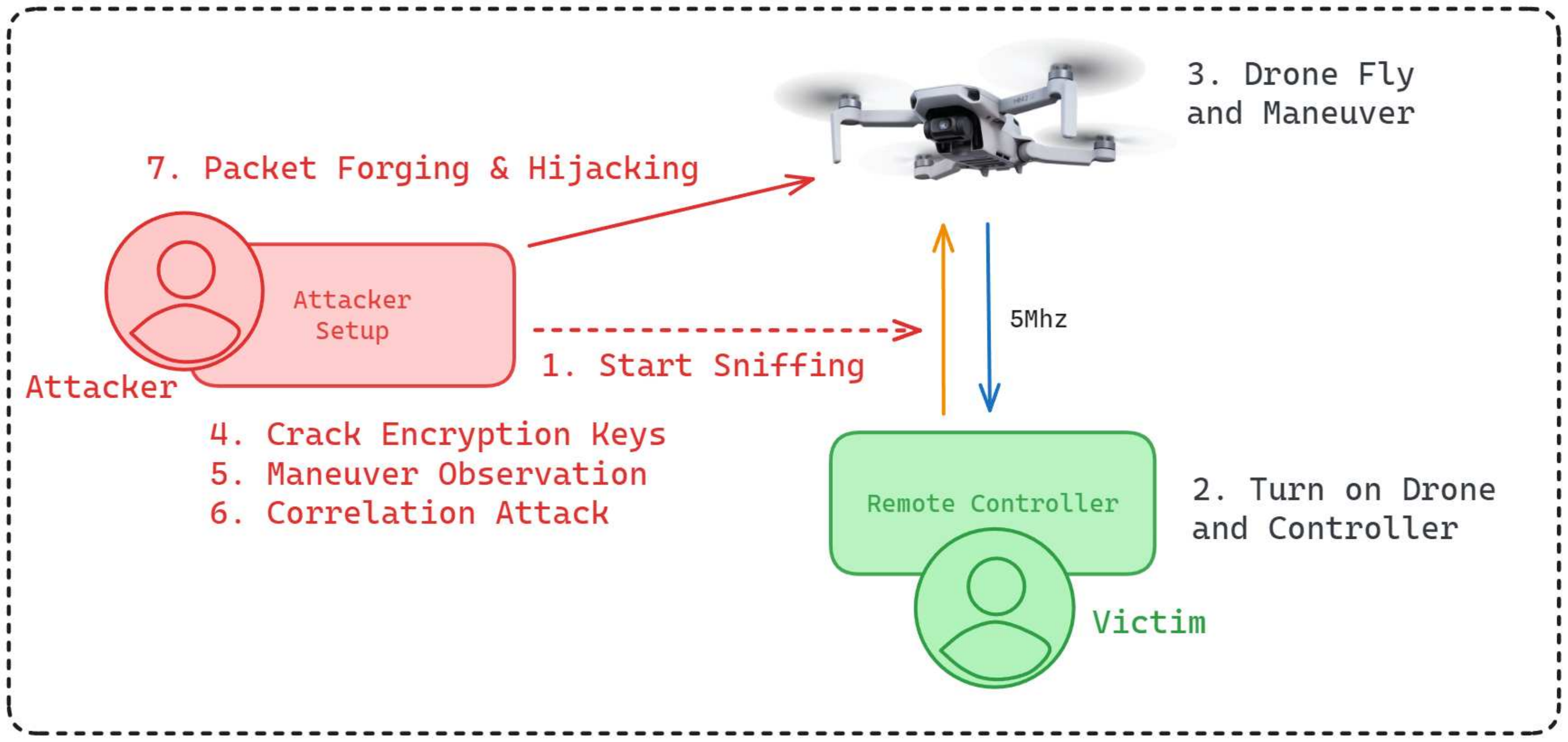}
    \centering
    \caption{Passive attack real-world scenario}
    \label{fig:threatmodel}
\end{figure}

Our research focuses on threat models where the attacker is within the effective range of the drone's communication capabilities. It involves a man-in-the-middle attack scenario, enabling the attacker to intercept and monitor the communication between the drone and its remote controller, shown in Figure~\ref{fig:threatmodel}. Below are the explanations:

\begin{enumerate}
  \item The attacker is able to start capturing the complete packet from drone initialization using a device that operates at the same protocol and frequency as the victim communication link, explained in Section~\ref{packetsniffing}.
  \item The victim turns on the drone and remote, allowing the attacker to sniff the encrypted initialization packets.
  \item The victim operates the drone and does some maneuvers.
  \item With enough IV from packets, the attacker is then able to crack the encryption in Section~\ref{section:wepcracking}.
  \item The attacker is able to observe the drone maneuver and its time.
  \item The attacker then deduces the specific control commands using a correlation analysis attack based on the drone maneuver observation as in Section~\ref{section:correlationanalysis}.
  \item The attacker can then forge a duplicate initialization and forge a new control command to take over the victim drone Section~\ref{section:controlhijackingattack}
\end{enumerate}

\begin{figure}[!h]
    \includegraphics[width= 0.40\textwidth] {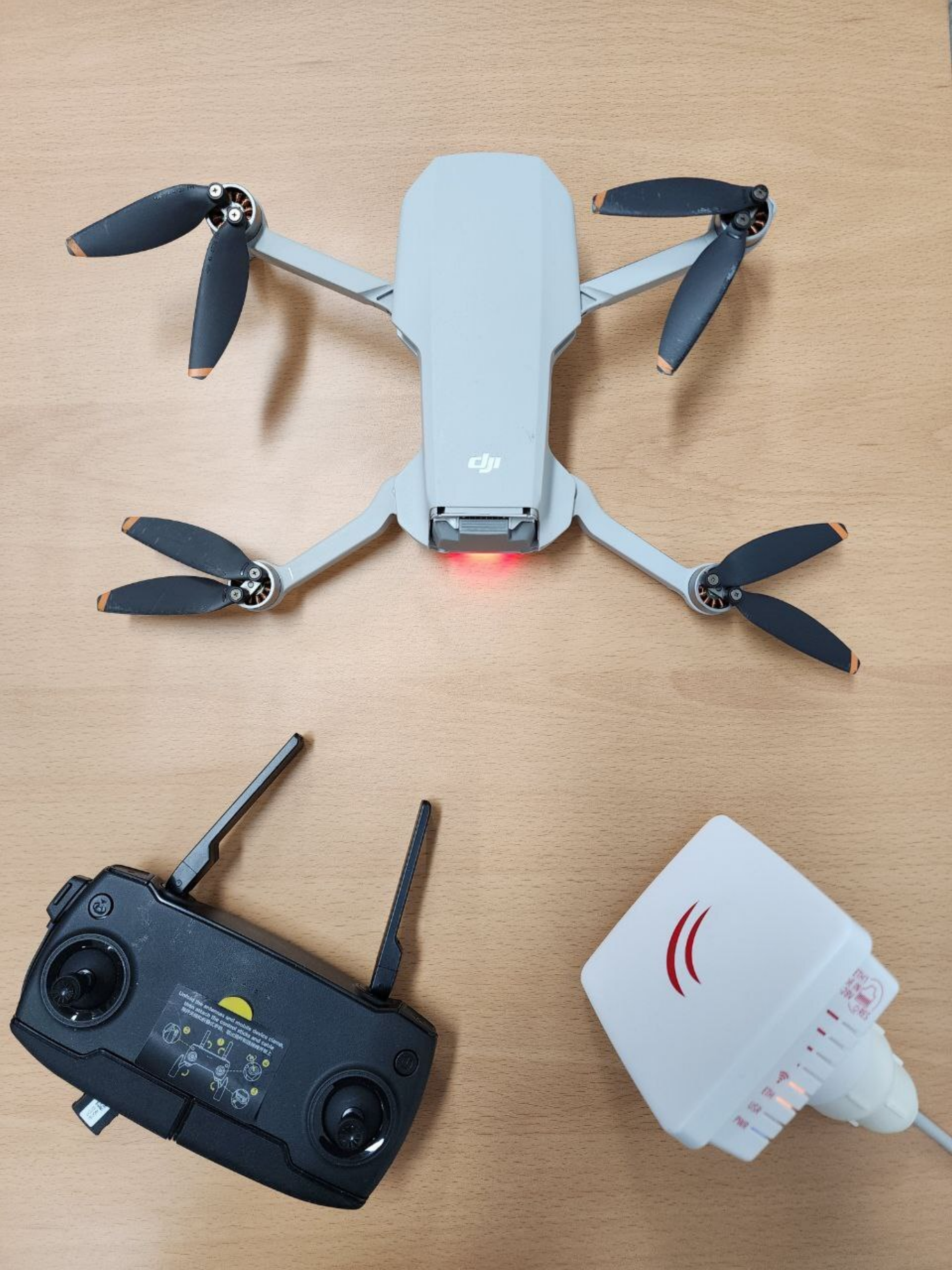}
    \centering
    \caption{Attack setup without any physical access to the target using only \$45 Wi-Fi router}
    \label{fig:realattacksetup}
\end{figure}

The drone can be in either a connected or disconnected state while being powered on. When the drone is disconnected, the LED turns red, as shown in Figure ~\ref{fig:realattacksetup}. As a target, we use a DJI Mini SE drone with the latest firmware version (01.02.0000) at the time of this research, which employs an Enhanced Wi-Fi communication protocol~\cite{ref_DJIMiniSpec}.

\subsection{Packet Sniffing\label{packetsniffing}}

DJI Mini SE drone uses Enhanced Wi-Fi protocol for remote and video transmission mentioned by the marketing page~\cite{ref_DJIMiniSpec}. There are two versions of it which support both only 5.8Ghz and both with 2.4Ghz. To simplify our sniffing process, we first turned off the automatic frequency selection on the drone application settings and set the channel to a 5.8GHz fixed channel at 149. Please note that this attack does not depend on static frequency settings, see Section \ref{freqhopping}.

\begin{figure}[h]
    \includegraphics[width=0.48\textwidth]{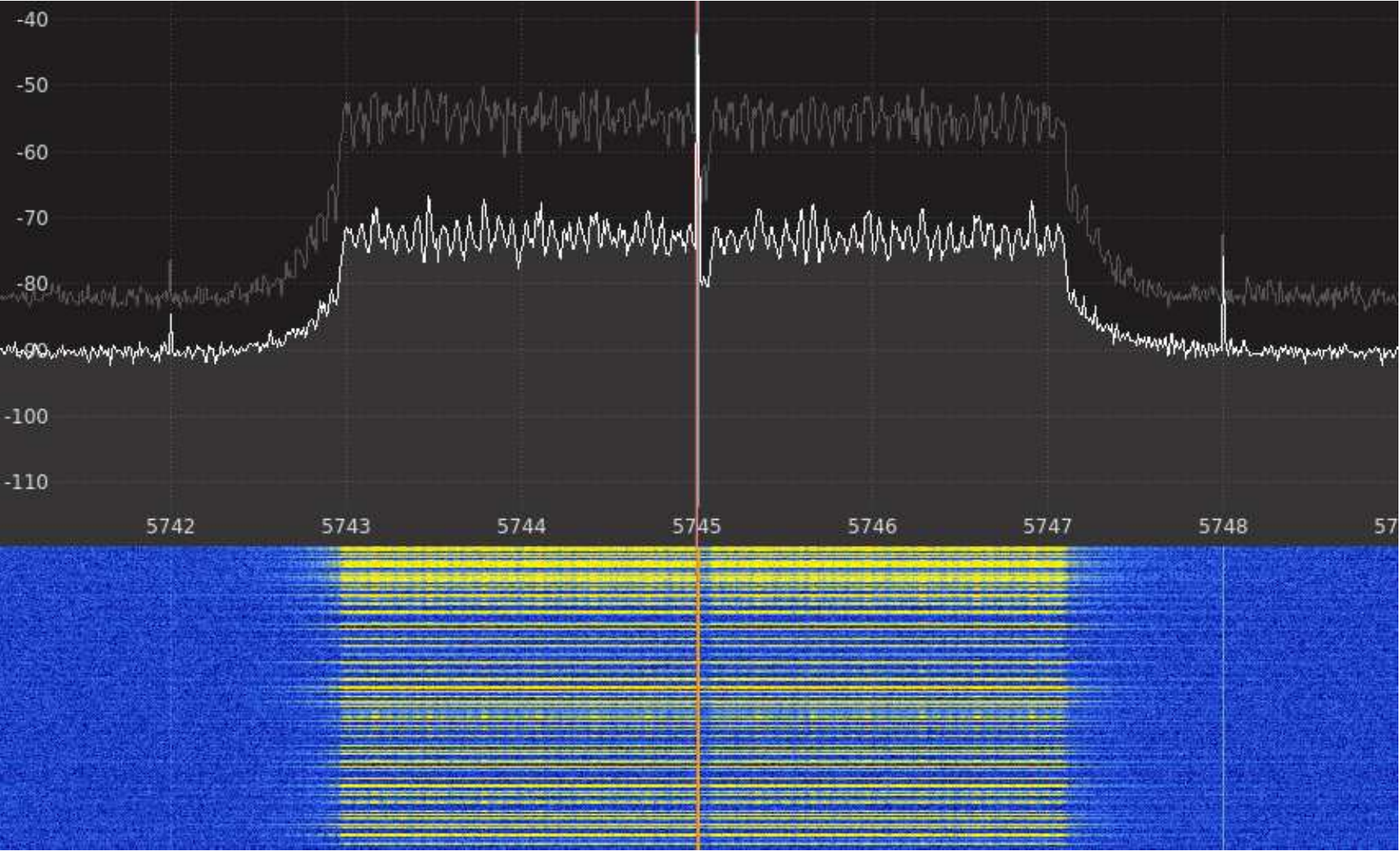}
    \centering
    \caption{HackRF shows that the width of the channel is only 5 Mhz, quarter half of the common bandwidth}
    \label{fig5mhzspectrum}
\end{figure}

In our examination, the ALFA Wi-Fi adapter AWUS036ACH, when utilized in monitor mode, failed to detect any indications of drone beacons. Consequently, we opted to employ Software Defined Radio (SDR), specifically the HackRF, which offers the capability to perform spectrum sweeps. Upon inspection of the communication spectrum at 5.745GHz using HackRF, we identified a signal exhibiting an atypical channel width of 5MHz, as delineated in the associated Figure ~\ref{fig5mhzspectrum}.

 We decode the signal using \texttt{gr-ieee802-11}~\cite{ref_Wi-Fi}, but the results were suboptimal, evidenced by missing packet-based observations on the sequence number, resulting in up to 63\% packet loss. Our hypothesis is that these deficiencies stem from hardware constraints, particularly when contrasting the performance between the high-end USRP Ettus N210 used in prior studies and the more economically priced SDR HackRF. The substantial disparity in both performance and cost between these devices is likely the root cause of the observed imperfections in the decoding process.

\begin{figure}[htbp]
    \centering
    \includegraphics[width=0.48\textwidth]{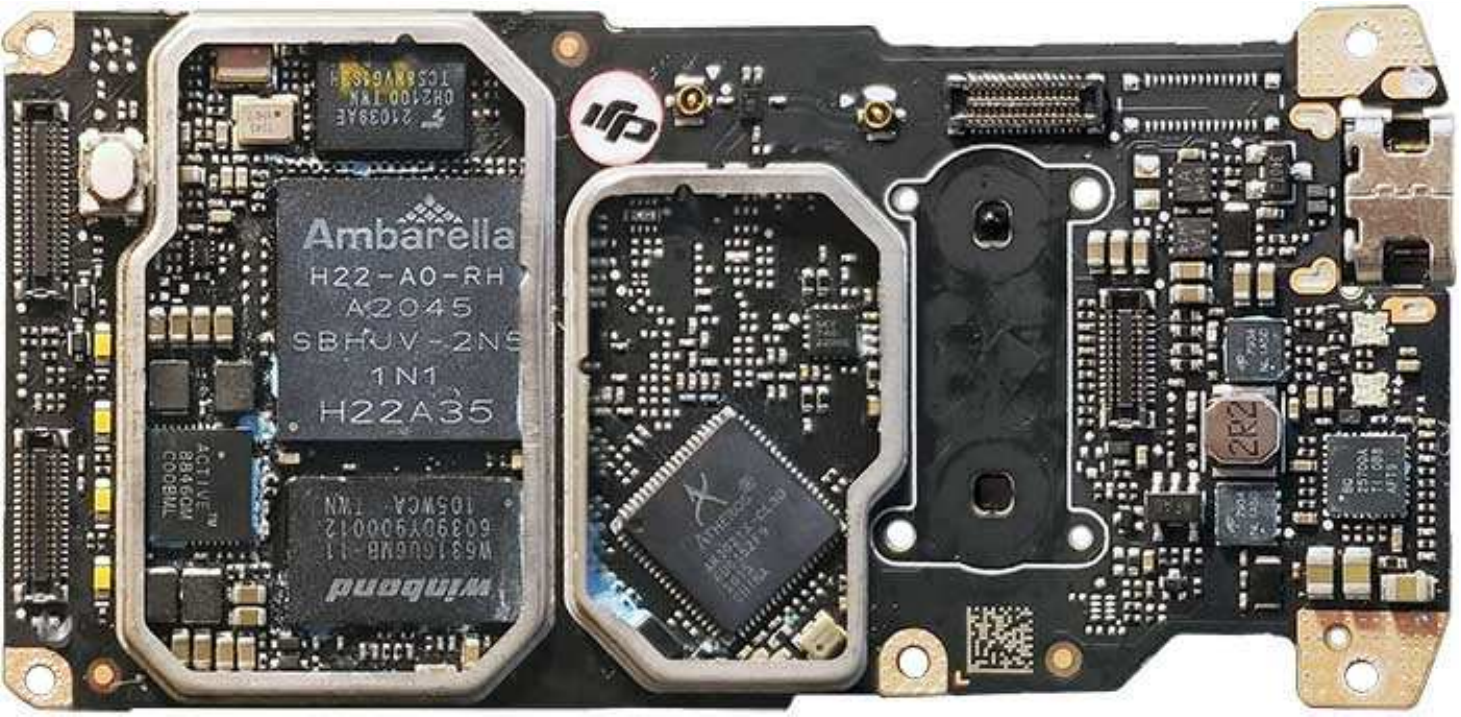}
    \caption{Main board shows the Atheros AR1021X-CL3D Wi-Fi module located in the middle of the board} \label{figmainboard}
\end{figure}

Utilizing open-source knowledge on Original Gangster’s (OG) repository~\cite{ref_DJIfirmware_git}, it appears that the drone is using Atheros AR1021X-CL3D as shown in Figure~\ref{figmainboard}. Therefore we investigate another 5MHz-supported Wi-Fi device. Most of them are no longer produced and outdated, but some are still in production and available for consumer use. Our search for specific devices that use Atheros 9K chips in the OpenWRT router database led us to Mikrotik LDF 5, a low-cost CPE router. We can sniff the communication between RC and the drone using Mikrotik LDF 5 at channel 149 with 5 Mhz channel bandwidth~\cite{ref_openwrt_toh}.

\begin{figure}[ht]
    \centering
    \includegraphics[width=0.48\textwidth]{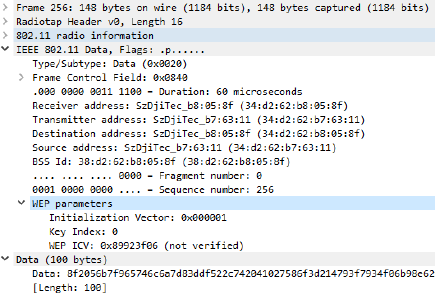}
    \caption{Packet shown the encryption being used is WEP} \label{figwepencrypted}
\end{figure}


In Figure~\ref{figwepencrypted}, we observed that the connection is established through an ad-hoc network utilizing Wired Equivalent Privacy (WEP) encryption. Given that WEP's vulnerabilities have been extensively documented and it is regarded as an insecure protocol, its usage in this context is unexpected and concerning. Notably, this discovery is all the more alarming considering that the target drone's firmware is at the most recent version (01.02.0000). This situation should ostensibly mitigate such outdated and insecure practices.


\section{Reverse Engineering}

\subsection{WEP Cracking\label{section:wepcracking}}

Utilizing the PTW (Pyshkin, Tews, Weinmann) method, we have successfully deciphered the key through the accumulation of sufficient IV (Initialization Vector) packets with the help of the tool \texttt{aircrack-ng}~\cite{ref_ePrint}. Notably, the key is static, remaining unchanged across various sessions during our data collection. Upon discovering the WEP key, we decrypted the communication, thereby revealing the uplink and downlink packets in plain text. An examination of the initial encrypted packet identified it as an ARP (Address Resolution Protocol) Request originating from the remote controller (RC).


\begin{figure*}[hb]
    \centering
    \includegraphics[width=\textwidth]{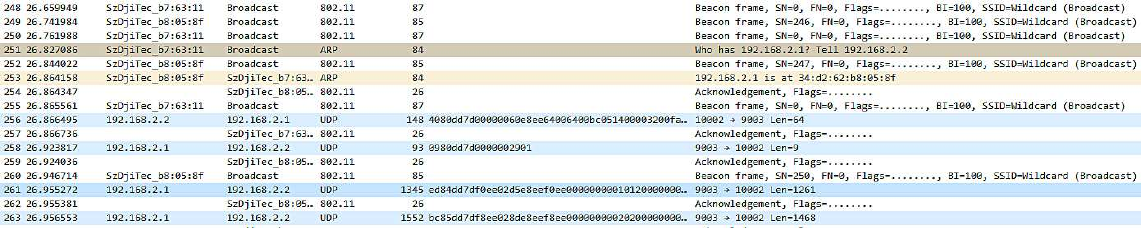}
    \caption{Sequence of packet shown after decryption} \label{figwepdecrypted}
\end{figure*}

In the initialization process, ARP requests are first utilized, followed by transmitting User Datagram Protocol (UDP) packets to the drone. Initially, the controller is interfaced with a mobile device, facilitating the relay of video data from the drone. However, for the purpose of our investigation, which is specifically concentrated on the analysis and evaluation of control command protocols, we establish a direct connection between the remote controller and the drone, devoid of the smartphone. This results in a simplified packet structure, thereby reducing complexity in the data communication pattern and aiding in our focused examination.



\subsection{Connection Authentication Analysis}

\begin{figure}[!h]
    \includegraphics[width=0.48\textwidth]{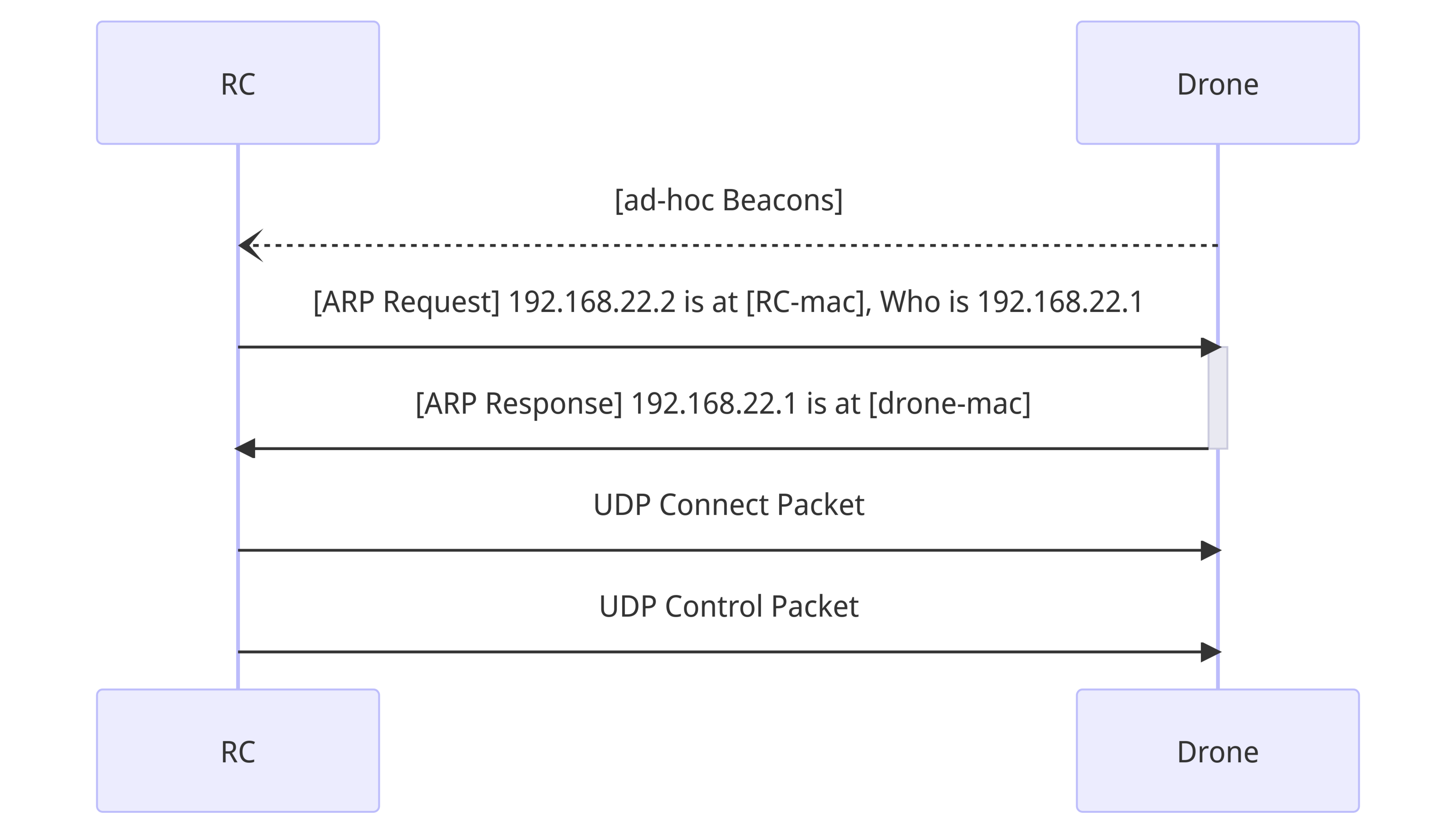}
    \centering
    \caption{Connection phase between the remote controller and drone}
    \label{rc-droneconnection}
\end{figure}

After we obtain the key to decrypt the connection, we record a full session and analyze the link initiation flow, as shown in Figure~\ref{rc-droneconnection}.

First, the drone turns on and sends a broadcast beacon packet ad hoc connection with a hidden SSID. Then the RC turns on and sends a broadcast beacon packet. RC then sent the first encrypted ARP request for 192.168.2.1. The drone response also encrypted the ARP response telling the RC about the MAC of the drone. Then RC sends the first connection UDP packet with WEP encryption. The drone sends replies with a short UDP packet which we call a connection initiator packet. Then the drone sends a few similar packets that contain control commands. In this phase, the drone is already connected to the controller.

    
\subsection{Frequency Hopping Detection\label{freqhopping}}
In the initial stages of our experiment, we operated using a static frequency to detect the beacon from the 'ad-hoc' type of connection. We discovered that even after activating the automatic frequency selection, we are still able to locate the beacon by scanning each frequency. This implies that the utilization of automatic frequency selection is not an effective countermeasure to prevent the attack methodology we have explored.

\subsection{Filtering}


By analyzing different session data, we observed a diversity of data lengths being transmitted. By employing Principal Component Analysis (PCA), we were able to identify the most commonly occurring data length, specifically \texttt{0x3C}, as outlined in Table~\ref{table:length}. We subsequently centered our analysis on the \texttt{0x3C} length of data to methodically investigate the individual bits responsible for controlling the drone's movements.
\begin{table}[h]
\caption{Occurrences of different packet lengths in one flight session}
\label{table:length}
\centering
\begin{tabular}{|c|c|c|c|c|c|c|c|c|c|}
\hline
\textbf{Length} & \textbf{40} & \textbf{3C} & \textbf{56} & \textbf{5D} & \textbf{22} & \textbf{A4} & \textbf{70} & \textbf{3D} & \textbf{57} \\
\hline
\textbf{Counts} & 4 & 1047 & 299 & 1 & 31 & 3 & 3 & 6 & 1 \\
\hline
\end{tabular}
\end{table}


We observed a consistent pattern in certain segments in our examination of the \texttt{0x3C} packet values. This observation aligns with prior research~\cite{ref_ReverseEng}, where the author identified the inclusion of DUML packets within the command structure. A group specializing in reverse engineering, known as Original Gangster (OG), has developed DUML dissector tools specific to these packets~\cite{ref_DJIfirmware_git}. These tools were instrumental in allowing us to pinpoint a 6-byte section within the control packets. Despite this progress, the precise functionality of the control command remains unknown. In an effort to further understand the structure, we analyzed the packet subsequent to the \texttt{0x55} DUML delimiter, disregarding the preceding segment.


\subsection{Correlation Analysis Attack\label{section:correlationanalysis}}
We employed a correlation analysis between the quantity of payload and the corresponding time reference, as defined in Equation \ref{eq:1}. This methodology enables us to infer the pertinent control mechanisms for each specific movement. Here, \(t_i\) represents the time in seconds, and \(c_i\) denotes the cumulative instances of observed control occurrences.

\begin{equation}[h] \label{eq:1}
r = \frac{{\sum_{i=0}^{n}(t_i - \bar{t})(c_i - \bar{c})}}{{\sqrt{{\sum_{i=0}^{n}(t_i - \bar{t})^2}\cdot{\sum_{i=0}^{n}(c_i - \bar{c})^2}}}}
\end{equation}



\begin{figure}[!h]
    \includegraphics[width=0.48\textwidth]{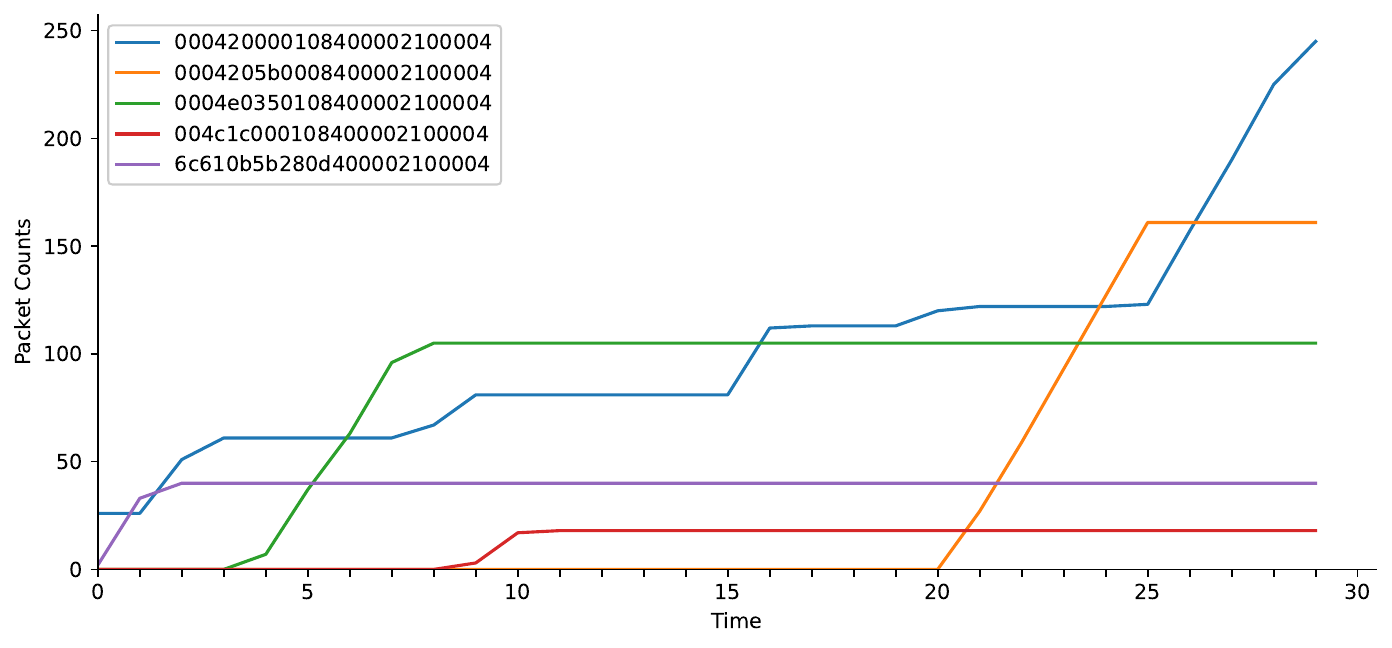}
    \centering
    \caption{Correlation graph when the drone flies backward then forward.}
    \label{correlationgraph}
\end{figure}

Figure~\ref{correlationgraph} shows an analysis of captured drone packet data corresponding to specific flight maneuvers, including take-off, a backward motion, a slight forward movement, and landing. It is imperative to recognize that an attacker can conduct a correlation analysis attack solely by observing the drone's aerial movement, thereby decoding the association between specific packets and individual movements.

The purple series demonstrates an initial increase, signifying packets responsible for activating the drone's propellers. Following this, the green series escalates at \( t = 3 \), corresponding to the take-off or "fly-up" packet. At \( t = 8 \), the drone moves backward, as evidenced by the red series packet. Though not explicitly captured here due to noise filtering, a slight forward motion is indicated, and the blue series (or "idle packet") remains static at \( t = 10 \), reflecting that the drone is in motion. The landing phase is marked at \( t = 20 \) in the orange series, with an observed increase in the idle packet immediately afterward.

This kind of data can lead to a more profound understanding of drone packet transmission and holds significant implications for security, particularly concerning potential unauthorized interception and manipulation of these controls.

Subsequently, we gather the entirety of the 6-byte control commands from all conditions for detailed analysis. We meticulously scrutinize each individual bit represented in Table~\ref{table_48bit}, organizing and identifying those which exhibit varying values and associated commands. The ones marked in bold represent static bits, which maintain a constant value across all commands.

\begin{table}[h]
\centering
    \caption{Bits associated with each command} 
    \label{table:associated}
    \begin{tabular}{|c|c|c|}
\hline
\textbf{Command}        & \textbf{\begin{tabular}[c]{@{}c@{}}Number of \\ Altered Bits\end{tabular}} & \textbf{Bit Position}          \\ \hline
Full Rotate Right       & 4 bits                                                            & 13, 11, 2, 0                         \\ \hline
Full Rotate Left        & 6 bits                                                            & 15, 14, 12, 11, {\bfseries 3}, 1     \\ \hline
Full Down               & 6 bits                                                            & 22, 20, 19, 17, 16, {\bfseries 8}    \\ \hline
Full Up                 & 4 bits                                                            & 23, 21, 18, 16                       \\ \hline
Full Forward            & 4 bits                                                            & 39, 37, 28, 26                       \\ \hline
Full Backward           & 6 bits                                                            & 38, 37, {\bfseries 29}, 27, 25, 24   \\ \hline
Full Fly Right          & 4 bits                                                            & 47, 44, 42, 33                       \\ \hline
Full Fly Left           & 6 bits                                                            & 46, 45, 43, 42, {\bfseries 34}, 3    \\ \hline
\begin{tabular}[c]{@{}c@{}}Ready\\(Propeller ON)\end{tabular} & 22 bits                     & \begin{tabular}[c]{@{}c@{}}46, 45, 43, 42, 38, 37, {\bfseries 34},\\ 32, {\bfseries 29}, 27, 25, 24, 22, 20,\\ 19, 17, 16, 13, 11, {\bfseries 8}, 2, 0\end{tabular} \\ \hline
\end{tabular}
\end{table}

We conclude the associated bits in Table~\ref{table:associated}, in which the bold number text indicates the bit is active low and the other is active high. Across all commands, excluding the Ready command, the number of altered bits is always four, but when there is an active low bit, the number is always 6. The active low bits (3, 8, 29, 34) always happen when the controller is in the left and down position, meaning that the active low bit is the negative value indicator. This is also proven with 3-bits (8, 29, 34) are active low when executing the Ready command, which moves the right joystick position between fly left and backward combined with the left joystick position between rotate right and down. Once we know the command control value, we can prepare to authenticate as controller and inject these values into the drone.

\begin{table*}[ht]
\centering
    \caption{The summary of 6 bytes of each command}
    \label{table_48bit}
    \begin{tabular}{|P{0.5cm}|P{0.26cm}|P{0.26cm}|P{0.26cm}|P{0.26cm}|P{0.26cm}|P{0.26cm}|P{0.26cm}|P{0.26cm}|P{0.26cm}|P{0.26cm}|P{0.26cm}|P{0.26cm}|P{0.26cm}|P{0.26cm}|P{0.26cm}|P{0.26cm}|P{0.26cm}|P{0.26cm}|P{0.26cm}|P{0.26cm}|P{0.26cm}|P{0.26cm}|P{0.26cm}|P{0.26cm}|P{0.26cm}|P{0.26cm}|P{0.26cm}|}
    \hline
    {\centering CMD} & 47 & 46 & 45 & 44 & 43 & 42 & {\bfseries 41} & {\bfseries 40} & 39 & 38 & 37 & {\bfseries 36} & {\bfseries 35} & 34 & 33 & 32 & {\bfseries 31} & {\bfseries 30} & 29 & 28 & 27 & 26 & 25 & 24 \\
    \hline
    {\centering I} & 0 & 0 & 0 & 0 & 0 & 0 & {\bfseries 0} & {\bfseries 0} & 0 & 0 & 0 & {\bfseries 0} & {\bfseries 0} & 1 & 0 & 0 & {\bfseries 0} & {\bfseries 0} & 1 & 0 & 0 & 0 & 0 & 0 \\
    \hline
    {\centering FRR} & 0 & 0 & 0 & 0 & 0 & 0 & {\bfseries 0} & {\bfseries 0} & 0 & 0 & 0 & {\bfseries 0} & {\bfseries 0} & 1 & 0 & 0 & {\bfseries 0} & {\bfseries 0} & 1 & 0 & 0 & 0 & 0 & 0 \\
    \hline
    {\centering FRL} & 0 & 0 & 0 & 0 & 0 & 0 & {\bfseries 0} & {\bfseries 0} & 0 & 0 & 0 & {\bfseries 0} & {\bfseries 0} & 1 & 0 & 0 & {\bfseries 0} & {\bfseries 0} & 1 & 0 & 0 & 0 & 0 & 0 \\
    \hline
    {\centering FD} & 0 & 0 & 0 & 0 & 0 & 0 & {\bfseries 0} & {\bfseries 0} & 0 & 0 & 0 & {\bfseries 0} & {\bfseries 0} & 1 & 0 & 0 & {\bfseries 0} & {\bfseries 0} & 1 & 0 & 0 & 0 & 0 & 0 \\
    \hline
    {\centering FU} & 0 & 0 & 0 & 0 & 0 & 0 & {\bfseries 0} & {\bfseries 0} & 0 & 0 & 0 & {\bfseries 0} & {\bfseries 0} & 1 & 0 & 0 & {\bfseries 0} & {\bfseries 0} & 1 & 0 & 0 & 0 & 0 & 0 \\
    \hline
    {\centering FFW} & 0 & 0 & 0 & 0 & 0 & 0 & {\bfseries 0} & {\bfseries 0} & 1 & 0 & 1 & {\bfseries 0} & {\bfseries 0} & 1 & 0 & 0 & {\bfseries 0} & {\bfseries 0} & 1 & 1 & 0 & 1 & 0 & 0 \\
    \hline
    {\centering FB} & 0 & 0 & 0 & 0 & 0 & 0 & {\bfseries 0} & {\bfseries 0} & 0 & 1 & 1 & {\bfseries 0} & {\bfseries 0} & 1 & 0 & 0 & {\bfseries 0} & {\bfseries 0} & 0 & 0 & 1 & 0 & 1 & 1 \\
    \hline
    {\centering FFR} & 1 & 0 & 0 & 1 & 0 & 1 & {\bfseries 0} & {\bfseries 0} & 0 & 0 & 0 & {\bfseries 0} & {\bfseries 0} & 1 & 1 & 0 & {\bfseries 0} & {\bfseries 0} & 1 & 0 & 0 & 0 & 0 & 0 \\
    \hline
    {\centering FFL} & 0 & 1 & 1 & 0 & 1 & 1 & {\bfseries 0} & {\bfseries 0} & 0 & 0 & 0 & {\bfseries 0} & {\bfseries 0} & 0 & 0 & 1 & {\bfseries 0} & {\bfseries 0} & 1 & 0 & 0 & 0 & 0 & 0 \\
    \hline
    {\centering RDY} & 0 & 1 & 1 & 0 & 1 & 1 & {\bfseries 0} & {\bfseries 0} & 0 & 1 & 1 & {\bfseries 0} & {\bfseries 0} & 0 & 0 & 1 & {\bfseries 0} & {\bfseries 0} & 0 & 0 & 1 & 0 & 1 & 1 \\
    \hline\hline
    CMD& 23 & 22 & 21 & 20 & 19 & 18 & 17 & 16 & 15 & 14 & 13 & 12 & 11 & {\bfseries 10} & {\bfseries 9} & 8 & {\bfseries 7} & {\bfseries 6} & {\bfseries 5} & {\bfseries 4} & 3 & 2 & 1 & 0 \\
    \hline
    {\centering I} & 0 & 0 & 0 & 0 & 0 & 0 & 0 & 0 & 0 & 0 & 0 & 0 & 0 & {\bfseries 0} & {\bfseries 0} & 1 & {\bfseries 0} & {\bfseries 0} & {\bfseries 0} & {\bfseries 0} & 1 & 0 & 0 & 0 \\
    \hline
    {\centering FRR} & 0 & 0 & 0 & 0 & 0 & 0 & 0 & 0 & 0 & 0 & 1 & 0 & 1 & {\bfseries 0} & {\bfseries 0} & 1 & {\bfseries 0} & {\bfseries 0} & {\bfseries 0} & {\bfseries 0} & 1 & 1 & 0 & 1 \\
    \hline
    {\centering FRL} & 0 & 0 & 0 & 0 & 0 & 0 & 0 & 0 & 1 & 1 & 0 & 1 & 1 & {\bfseries 0} & {\bfseries 0} & 1 & {\bfseries 0} & {\bfseries 0} & {\bfseries 0} & {\bfseries 0} & 0 & 0 & 1 & 0 \\
    \hline
    {\centering FD} & 0 & 1 & 0 & 1 & 1 & 0 & 1 & 1 & 0 & 0 & 0 & 0 & 0 & {\bfseries 0} & {\bfseries 0} & 0 & {\bfseries 0} & {\bfseries 0} & {\bfseries 0} & {\bfseries 0} & 1 & 0 & 0 & 0 \\
    \hline
    {\centering FU} & 1 & 0 & 1 & 0 & 0 & 1 & 0 & 1 & 0 & 0 & 0 & 0 & 0 & {\bfseries 0} & {\bfseries 0} & 1 & {\bfseries 0} & {\bfseries 0} & {\bfseries 0} & {\bfseries 0} & 1 & 0 & 0 & 0 \\
    \hline
    {\centering FFW} & 0 & 0 & 0 & 0 & 0 & 0 & 0 & 0 & 0 & 0 & 0 & 0 & 0 & {\bfseries 0} & {\bfseries 0} & 1 & {\bfseries 0} & {\bfseries 0} & {\bfseries 0} & {\bfseries 0} & 1 & 0 & 0 & 0 \\
    \hline
    {\centering FB} & 0 & 0 & 0 & 0 & 0 & 0 & 0 & 0 & 0 & 0 & 0 & 0 & 0 & {\bfseries 0} & {\bfseries 0} & 1 & {\bfseries 0} & {\bfseries 0} & {\bfseries 0} & {\bfseries 0} & 1 & 0 & 0 & 0 \\
    \hline
    {\centering FFR} & 0 & 0 & 0 & 0 & 0 & 0 & 0 & 0 & 0 & 0 & 0 & 0 & 0 & {\bfseries 0} & {\bfseries 0} & 1 & {\bfseries 0} & {\bfseries 0} & {\bfseries 0} & {\bfseries 0} & 1 & 0 & 0 & 0 \\
    \hline
    {\centering FFL} & 0 & 0 & 0 & 0 & 0 & 0 & 0 & 0 & 0 & 0 & 0 & 0 & 0 & {\bfseries 0} & {\bfseries 0} & 1 & {\bfseries 0} & {\bfseries 0} & {\bfseries 0} & {\bfseries 0} & 1 & 0 & 0 & 0 \\
    \hline
    {\centering RDY} & 0 & 1 & 0 & 1 & 1 & 0 & 1 & 1 & 0 & 0 & 1 & 0 & 1 & {\bfseries 0} & {\bfseries 0} & 0 & {\bfseries 0} & {\bfseries 0} & {\bfseries 0} & {\bfseries 0} & 1 & 1 & 0 & 1 \\
    \hline
    \end{tabular}
    \begin{tablenotes}
      \small
      \item {\bfseries *Note} CMD: Command, I: Idle Condition; FRR: Full Rotate Right; FRL: Full Rotate Left; FD: Full Down; FU: Full Up; FFW: Full Forward; FB: Full Backward; FFR: Full Fly Right; FFL: Full Fly Left; RDY: Propeller On. {\bfseries Bold}: Static across all commands.
    \end{tablenotes}
\end{table*}

\section{Hijacking Evaluation}

\subsection{Analog Replay Attack}
A key flaw within WEP's design is its lack of protection against replay attacks. By leveraging this weakness, an attacker can record the analog signal wave corresponding to the connection initiation process using tools such as HackRF. Subsequently, the recorded signal can be replayed to the drone, enabling the attacker to seize control in the event of a disconnection from the legitimate owner. We are also able to inject the recorded Wi-Fi packets into the drone using monitor mode Mikrotik LDF 5, which costs only 18\% of the HackRF price.
\subsection{Control Hijacking Attack\label{section:controlhijackingattack}}

\subsubsection{Packet Forging and Injection}
This research aims to manipulate the control of the drone by command. We reverse engineer the control command instead of just attempting a replay attack.




\begin{figure}[h]
    \includegraphics[width=0.48\textwidth]{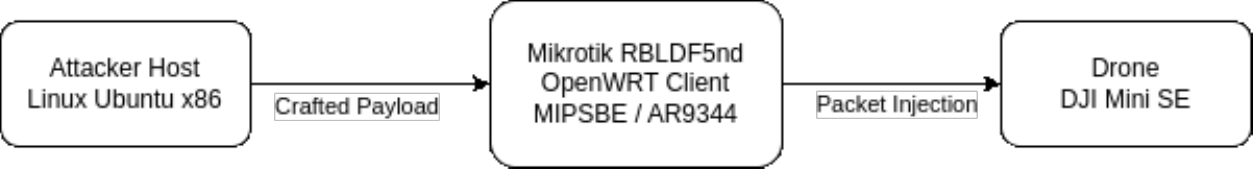}
    \centering
    \caption{Hijacking setup}
    \label{fig:attackersetup}
\end{figure}

We forge the packet with Scapy~\ref{scapy}, which helps us easily modify and generate 802.11 packets. At first, we craft only the ARP and UDP packet, which is encrypted using WEP. However, we found that initiating a connection is not enough. Based on the observation, another packet other than UDP keeps showing.

There are acknowledge and beacon packets. The connection needs to be maintained by sending the controller beacon frame continuously. Surprisingly enough, we don't need to send acknowledgment packets to maintain a connection. We assumed that the Beacon packets from the controller were a sign to a controller that they were using the same channel.

\begin{figure}[h]
    \includegraphics[width=0.48\textwidth]{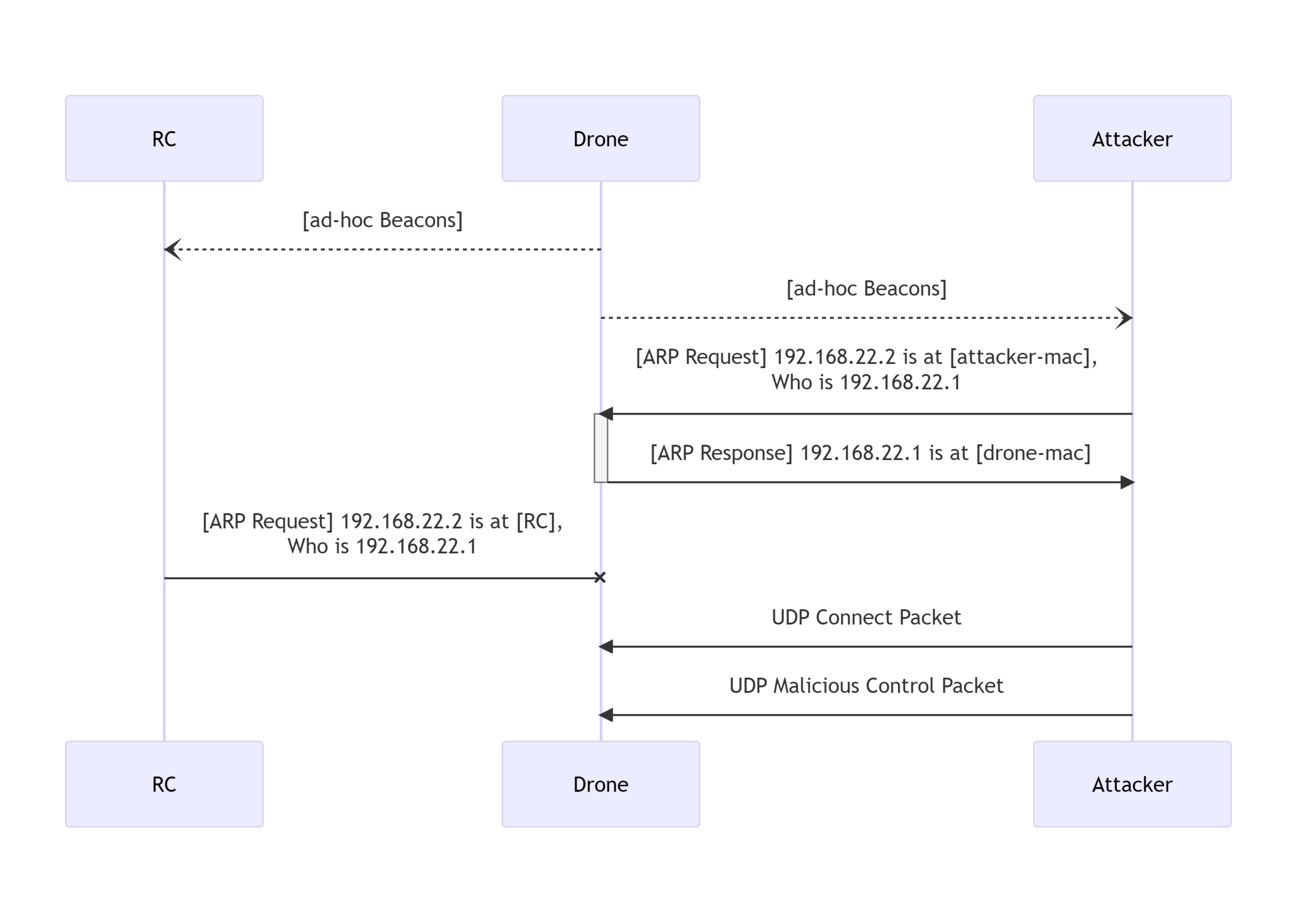}
    \centering
    \caption{Connection hijacking phase}
    \label{connectionhijack}
\end{figure}

For the purpose of commandeering the drone, we identified three fundamental packet formats to be employed: the beacon, an ARP request, and the UDP packets:

\begin{itemize}
  \item Beacon packet intervals are needed to inform the connection speed, and we maintain this in the background.
  \item ARP request packet is needed to lure the drone into telling its MAC address.
  \item At the same time,  after the ARP request is sent, the drone will accept a connection initiator packet, the captured \texttt{0x40} length UDP packet.
\end{itemize}

These three packets are used as captured without further modification. However, for the UDP control command packet, after we calculate the movement value, we have to recalculate the CRC value in Table~\ref{table:control_pkt_struct}. We use Kermit CRC with seed value from OG’s repository~\cite{ref_DJIfirmware_git}.


\begin{table}[h]
\caption{Structure of \texttt{0x3C} Length UDP Control Packet}\label{table:control_pkt_struct}
\centering
\begin{tabular}{p{0.5cm}>{\raggedleft\arraybackslash}p{0.5cm}p{0.5cm}>{\raggedleft\arraybackslash}p{1cm}p{0.5cm}>{\raggedleft\arraybackslash}p{1cm}p{0.5cm}>{\raggedleft\arraybackslash}p{0.5cm}}
0 & 45 & 46 & 51 & 52 & 57 & 58 & 59 \\ \hline
\multicolumn{2}{|c|}{Prefix} & \multicolumn{2}{c|}{Movement} & \multicolumn{2}{c|}{Unknown} & \multicolumn{2}{c|}{CRC Kermit} \\ \hline
\multicolumn{2}{c}{46} & \multicolumn{2}{c}{6} & \multicolumn{2}{c}{6} & \multicolumn{2}{c}{2}
\end{tabular}
\end{table}




We then inject the crafted packets. We inject the command using HackRF and Mikrotik LDF 5.
Using HackRF, we implemented Wi-Fi TX modulation from the previous research, a widely known IEEE802-11 library for GnuRadio, the \texttt{gr-ieee802-11}~\cite{ref_Wi-Fi}, but the TX packet is sometimes lost and unstable, and the signal is also weak. This is the same problem as we mentioned before in Section \ref{packetsniffing} where the amounts of packet loss rate is more than 50\%. 

\begin{figure}[!h]
    \includegraphics[width=0.40\textwidth]{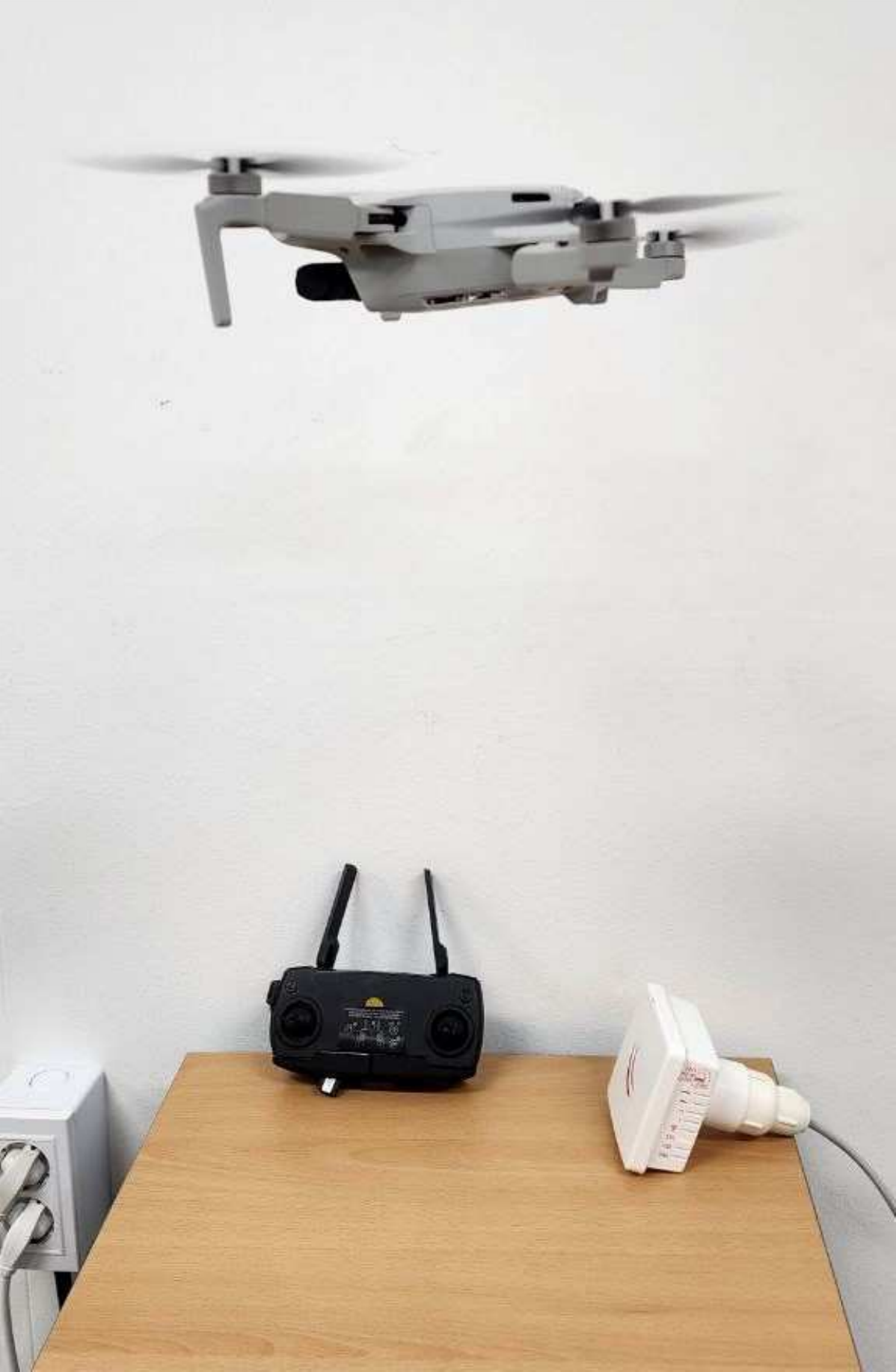}
    \centering
    \caption{Hijacked drone with disconnected controller, control packets are fully sent from the Wi-Fi router}
    \label{fig:drone-hijacked}
\end{figure}

We then switch to Mikrotik LDF 5, which already uses OpenWRT as the operating system. However, since Mikrotik LDF 5 has storage constraints, we are unable to install Scapy to craft the packet. An open-source program called etherpuppet~\cite{ref_etherpuppet} was available to utilize the constrained Wi-Fi device as a client to send the crafted packets by the host as a server using Scapy. However, \texttt{etherpuppet} does not provide the ability to send raw ethernet packets. Also, there is no way to detect the current Wi-Fi interface when operating in monitor mode. Therefore we developed our version of etherpuppet~\cite{ref_etherpuppet}, which will be used to send crafted packets on monitor mode Wi-Fi interface. The attacker host forges the packet and then sends it via TCP in the ethernet interface to the Wi-Fi device, which then resends it in a 5Mhz channel bandwidth as seen in Figure ~\ref{fig:attackersetup}. In this way, we are able to hijack the drone and control it with all of our commands from Table~\ref{table_48bit} as seen in Figure ~\ref{fig:drone-hijacked}.

In Table~\ref{table:hijackresult}, we tested each of all the commands we deduced from the correlation analysis attack in~\ref{section:correlationanalysis}. We are successfully able to hijack the drone control regardless of the connection state of the legitimate controller. Whether the legitimate controller is connected or disconnected, the drone recognizes and accepts our crafted packets as legitimate controller packets.

\begin{table*}[!hb]
\caption{Hijack control test result}
\label{table:hijackresult}
\centering
\begin{tabular}{|c|c|c|c|c|c|c|c|c|c|c|}
\hline
\textbf{Control}  & Idle  & Propeller & Rotate Right & Rotate Left & Down  & Up    & Forward & Backward & Rightward & Leftward \\ \hline
\textbf{Verified} & \cmark & \cmark     & \cmark        & \cmark       & \cmark & \cmark & \cmark   & \cmark    & \cmark     & \cmark    \\ \hline
\end{tabular}
\end{table*}




\subsubsection{Key of Connection}
Based on our experiments, we found that two ARP requests can be sent without affecting the drone's response. Both the RC and our device were able to connect and control the drone simultaneously. The drone doesn't require an acknowledgment packet and still executes commands even if it's not received. The RC Beacon packet is important for checking communication speed, and sending idle packets after the connection initiator is necessary to maintain the connection. The drone can handle multiple simultaneous connection initiations and accept controls from legitimate RC devices and potential attackers. The HackRF device has a weaker signal compared to the original RC, while the Wi-Fi router injection performs better.

\section{Discussion}
We discuss the countermeasure for this kind of attack and the current security used by DJI. Also, we compare this drone's vulnerability with another higher-level drone.

\subsection{Ethical Disclosure}
On June 2, 2023, we responsibly disclosed a specific vulnerability to DJI Security, outlining the details of the potential security flaw. Acknowledging our initial communication, we followed up on June 12, 2023, with the full proof-of-concept (PoC) code, proving the potential attack vector.

DJI promptly responded on June 13, stating that they had analyzed the attack and promising to reply soon with their findings. Upon further inquiry, DJI informed us that the issue was already a known security vulnerability within their system. However, despite multiple attempts to ascertain the timeline for a fix or further details, DJI remained unresponsive.

Given the significant nature of this vulnerability and the lack of engagement from DJI to address the issue in a timely manner, we feel compelled to open-source the proof-of-concept on Github repository\footnote{Github Repository: \href{https://github.com/ibndias/dji-drone-hijacking}{https://github.com/ibndias/dji-drone-hijacking}}. This decision is in alignment with the principles of responsible disclosure and our commitment to ensuring the broader community is aware of potential risks, allowing others to take appropriate measures to mitigate potential threats.
\subsection{Countermeasure}
The AR1021X Wi-Fi module is natively compatible with WPA2 encryption, though the utilization of WEP in certain contexts is puzzling. To thwart potential attacks, two primary defenses can be employed. Firstly, the implementation of a more robust encryption protocol such as WPA2 is advisable. Despite known susceptibilities to Krack attacks as referenced in ~\cite{ref_kracken2}, WPA2 nonetheless presents a more formidable challenge to potential attackers, thereby increasing the complexity of any attempted intrusion or decryption of communications.

We can implement an intrusion detection system into the connection or block any other ARP spoofing packet once the genuine ARP Request from RC is made. This will prevent the attacker from sending commands with a different MAC than the controller. However, if the attacker uses the same MAC address, then it needs a more advanced detection system. Based on all of the above, the best countermeasure is to update the firmware and use higher security encryption for the Wi-Fi encryption.


\subsection{Ocusync}
In the Mini 3 Pro, the architecture of the communication chip diverges significantly from its predecessor, the Mini 2. The former employs a P1 SoC (Pigeon) while the latter utilizes an S1 SoC (Sparrow) ASIC chip, which encompasses both ARM and SDR. This distinction manifests as a 0.5 MHz offset in the connection. Although we have identified a router operating at this frequency, it is essential to note that the underlying communication protocol is not based on conventional Wi-Fi standards. In previous research, as outlined in~\cite{ref_droneid}, evidence suggests that the DroneID can be detected and unencrypted. This finding enhances the probability that, given the appropriate equipment for sniffing the Ocusync technology, the associated communication could potentially be decrypted, decoded, and thus made susceptible to the injection of malicious packets. This revelation opens a new frontier in our understanding of the security protocols surrounding these devices and mandates further investigation into the potential vulnerabilities they may conceal.

\subsection{Other Wi-Fi Based Drones}
The methodology employed in our correlation analysis attack can be extended to various drone systems, enabling the disclosure of specific control commands corresponding to distinct maneuvers, provided the adversary is observing the targeted drones. This assumes the attacker can decipher the encrypted communications between the drone and its control station. In scenarios where WPA2 encryption is utilized, the adversary may need to construct a rogue access point to facilitate a man-in-the-middle attack, thereby enabling the manipulation and interception of communication data.

\section{Conclusion}
In this paper, we reverse-engineer the control command of the Enhanced Wi-Fi protocol used in DJI Mini SE. We are able to hijack the drone whether in a currently controlled state or disconnected from the legitimate remote controller. We also perform security analysis on how the DJI control works. We show that the current Enhanced Wi-Fi protocol in DJI Drone is not secure enough and can be hijacked with only an off-the-self router. This paper proves that even the current market leader for civilian drones does not secure their drone correctly. Future civilian drones should use more secure communication protocols since the drone usage is already leveraged out there to be used as war weapons.


\ifCLASSOPTIONcompsoc
  \section*{Acknowledgments}
\else
  \section*{Acknowledgment}
\fi

The authors would like to thank...

\end{document}